\documentclass[prl,twocolumn,10pt,aps]{revtex4-1}

 \usepackage[utf8]{inputenc}
 \usepackage[greek,english]{babel}
 \usepackage{bm}
 \usepackage{verbatim}
 \usepackage{amsmath}
 \usepackage{amssymb}
 \usepackage{amsthm}
 \usepackage{latexsym}
 \usepackage{amsfonts}
 \usepackage{epsfig}
 \usepackage{epstopdf}
 \usepackage{wasysym} % for certain special symbols
 \usepackage{subfigure}
 \usepackage{color}
 \definecolor{darkblue}{rgb}{0,0,.5}
 \usepackage[linktocpage, colorlinks=true, linkcolor=darkblue, citecolor=darkblue]{hyperref}
 \usepackage[all]{hypcap}
 \usepackage[makeroom]{cancel}

\newcommand{\C}[1]{{\cal{#1}}}
\newcommand{\bb}[1]{\textbf{#1}}

\newcommand{\rl}[0]{{\rangle\langle}}

\def\dbar{{\mathchar'26\mkern-12mu d}}

\begin{document}

\title{Clausius Inequality for Finite Baths Reveals Universal Efficiency Improvements}

\author{Philipp Strasberg}
\author{Mar\'ia Garc\'ia D\'iaz}
\author{Andreu Riera-Campeny}
%\author{Anna Sanpera$^{1,2}$}
%\email{philipp.strasberg@uab.cat}
\affiliation{F\'isica Te\`orica: Informaci\'o i Fen\`omens Qu\`antics, Departament de F\'isica, Universitat Aut\`onoma de Barcelona, 08193 Bellaterra (Barcelona), Spain}
%\affiliation{$^2$ICREA -- Instituci\'o Catalana de Recerca i Estudis Avan\c{c}ats, Pg.~Lluis Companys, 23, 08010 Barcelona, Spain}

\date{\today}

\begin{abstract}
 We study entropy production in nanoscale devices, which are coupled to finite heat baths. This situation is of 
 growing experimental relevance, but most theoretical approaches rely on a formulation of the second law valid only 
 for infinite baths. We fix this problem by pointing out that already Clausius' paper from 1865 contains an adequate 
 formulation of the second law for finite heat baths, which can be also rigorously derived from a microscopic quantum 
 description. This Clausius' inequality shows that nonequilibrium processes are less irreversible than previously 
 thought. We use it to correctly extend Landauer's principle to finite baths and we demonstrate that any heat engine 
 in contact with finite baths has a higher efficiency than previously thought. Importantly, our results are easy to 
 study, requiring only the knowledge of the average bath energy. 
\end{abstract}

\maketitle
%\tableofcontents

%\newtheorem{mydef}{Definition}[section]
\newtheorem{lemma}{Lemma}[section]
%\newtheorem{thm}{Theorem}[section]
%\newtheorem{crllr}{Corollary}[section]
%\newtheorem*{thm*}{Theorem}%[section]
%\theoremstyle{remark}
%\newtheorem{rmrk}{Remark}[section]

%%%%%%%%%%%%%%%%%%%%%%%%%%%%%%%%%%%%%%%%%%%%%%%%%%%%%%%%%%%%%%%%%%%%%%%%%%%%%%%%%%%%%%%%%%%%%%%%%%%%%%%%%%%%%%%%%%%%%%%%
%\emph{Phenomenology.---}
More than 150 years ago, Clausius wrote down the following inequality, which now bears 
his name~\cite{Clausius1865}: 
\begin{equation}\label{eq Clausius}
 \Sigma \equiv \Delta S_S(\tau) - \int \frac{\dbar Q(t)}{T(t)} \ge 0.
\end{equation}
For that inequality to be valid, Clausius imagined a process where a system $S$ undergoes a nonequilibrium 
transformation for a time $\tau$ while being in contact with a heat bath with a time-dependent temperature $T(t)$. 
The change in thermodynamic entropy of the system is $\Delta S_S(\tau) = S_S(\tau) - S_S(0)$ and the 
infinitesimal heat flux from the bath \emph{into} the system at time $t$ is $\dbar Q(t)$. An ideal heat bath is 
characterized by the relation $dS_B(t) = -\dbar Q(t)/T(t)$, where $dS_B(t)$ is the infinitesimal change in bath 
entropy. Consequently, Eq.~(\ref{eq Clausius}) coincides with the traditional statement of the second law: 
the thermodynamic entropy of the universe (i.e., the system \emph{and} the bath) can not decrease. Hence, 
Eq.~(\ref{eq Clausius}) is often called the \emph{entropy production}, which we denote by $\Sigma$. 

If the bath temperature $T(t) = T(0)$ stays constant in time, which is the case if the duration 
$\tau$ is short or the bath large enough, Clausius' inequality reduces to  
\begin{equation}\label{eq Clausius 2}
 \Sigma' \equiv \Delta S_S(\tau) - \frac{Q(\tau)}{T(0)} \ge 0.
\end{equation}
Here, $Q(\tau) = \int \dbar Q(t)$ denotes the total flux of heat into the system. 
So far, our considerations were purely based on \emph{phenomenological} nonequilibrium thermodynamics. 

A central goal of statistical mechanics, in particular in the emerging field of quantum thermodynamics, is to 
explain the laws of thermodynamics based on underlying (reversible) quantum dynamics. Interestingly, 
the standard framework of quantum thermodynamics considers Eq.~(\ref{eq Clausius 2}) as the second law 
and Clausius' inequality~\cite{BinderEtAlBook2018, LandiPaternostroArXiv2020}---despite a rapidly growing interest in 
finite-sized heat baths~\footnote{In the entire book~\cite{BinderEtAlBook2018} Eq.~(\ref{eq Clausius}) is mentioned 
\emph{only once} in an introduction on page 681 and the recent review~\cite{LandiPaternostroArXiv2020} does not mention 
it either (see also references therein). }. 
Indeed, many interesting recent experiments operate far from the thermodynamic limit and finite size effects of the 
bath become visible~\cite{TrotzkyEtAlNP2012, BrantutEtAlScience2012, GringEtAlScience2012, 
BrantutEtAlScience2013, ClosEtAlPRL2016, KaufmanEtAlScience2016, KrinnerEsslingerBrantutJP2017, KarimiEtAlNC2020, 
BohlenEtAlPRL2020}, which calls for an urgent microscopic clarification of the relation between $\Sigma$ and 
$\Sigma'$. We here provide an information-theoretic identity [Eq.~(\ref{eq central}) below] demonstrating that 
$\Sigma' - \Sigma \ge 0$ always. To the best of our knowledge, this relation is not even known phenomenologically, 
and, as we show, it has profound consequences. 

To set the stage, we recall the by now well-known microscopic derivation of 
Eq.~(\ref{eq Clausius 2})~\cite{LindbladBook1983, PeresBook2002, EspositoLindenbergVandenBroeckNJP2010, 
SagawaUedaPRL2010b, TakaraHasegawaDriebePLA2010}. 
Consider a system coupled to a bath described by a joint quantum state $\rho_{SB}(t)$ evolving unitarily 
in time. Initially, the system is assumed decorrelated from a bath described by a canonical ensemble at temperature 
$T(0)$, which we denote by $\pi_B[T(0)] \equiv e^{-H_B/T(0)}/\C Z_B[T(0)]$. Here, $H_B$ is the bath Hamiltonian and 
$\C Z_B(T) \equiv \mbox{tr}_B\{e^{-H_B/T}\}$ the partition function ($k_B \equiv 1$). Now, solely based on the 
assumption that $\rho_{SB}(0) = \rho_S(0)\otimes\pi_B[T(0)]$, one can derive Eq.~(\ref{eq Clausius 2}) if one makes 
the following two identifications~\cite{LindbladBook1983, PeresBook2002, EspositoLindenbergVandenBroeckNJP2010, 
SagawaUedaPRL2010b, TakaraHasegawaDriebePLA2010}. First, the thermodynamic entropy of the \emph{system} is identified 
with the von Neumann entropy, $S_S(t) \equiv S_\text{vN}[\rho_S(t)]$, with 
$S_\text{vN}(\rho) \equiv -\mbox{tr}\{\rho\ln\rho\}$. Second, the heat flux into the system is identified 
with minus the change in bath energy, $Q(t) \equiv -\Delta E_B(t)$, with $E_B(t) \equiv \mbox{tr}_B\{H_B\rho_B(t)\}$. 

This derivation of Eq.~(\ref{eq Clausius 2}) is remarkable because \emph{no} assumption about the dynamics and the 
bath size enters it. However, for a finite bath it has not been possible to link the term $-Q(\tau)/T(0)$ to an 
entropy change. Hence, while remaining a valid mathematical inequality, Eq.~(\ref{eq Clausius 2}) is no longer 
identical to the second law. 

In this paper, we advocate the use of Eq.~(\ref{eq Clausius}) instead of Eq.~(\ref{eq Clausius 2}) for finite baths. 
In fact, very recently it was shown that---under the \emph{same} conditions as above---also Eq.~(\ref{eq Clausius}) 
can be microscopically derived~\cite{StrasbergWinterArXiv, RieraCampenySanperaStrasbergPRXQ2021}. The time-dependent 
bath temperature $T(t)$ appearing in this microscopic derivation is then fixed by demanding that the bath energy 
$E_B(t)$ matches the one computed with a canonical ensemble at temperature $T(t)$: 
\begin{equation}\label{eq def temperature}
 E_B(t) \overset{!}{=} \mbox{tr}\{H_B\pi_B[T(t)]\}.
\end{equation}
Note that we are \emph{not} asserting that $\rho_B(t) = \pi_B[T(t)]$, we are only using the given information 
$E_B(t)$ in the least biased (maximum entropy) way to infer $T(t)$. Microscopically, Eq.~(\ref{eq Clausius}) remains 
valid even if $\rho_B(t)$ is \emph{far} from 
\emph{canonical} equilibrium. Note that the phenomenological Clausius inequality does not rely either 
on a canonical ensemble. In fact, typicality arguments and the eigenstate thermalization hypothesis have convincingly 
shown that many quantum states have a well defined macroscopic temperature $T$, even pure states `far' from 
$\pi_B(T)$~\cite{DeutschRPP2018}. Thus, if the equivalence of ensembles holds~\cite{TouchetteJSP2015}, our microscopic 
description matches well the phenomenological theory. In general, of course, the correct definition of a 
nonequilibrium temperature is a complicated open question beyond the present scope~\cite{CasasVazquezJouRPP2003}. 

Our approach based on a microscopically emerging definition of temperature also generalizes the few notable 
previous studies, which considered a time-dependent bath temperature. They either assumed that $T(t)$ is externally 
controllable in time~\cite{JarzynskiJSP1999, BrandnerSaitoSeifertPRX2015, BrandnerSeifertPRE2016, BrandnerSaitoPRL2020} 
or $T(t)$ was dynamically determined in the linear response regime for baths that do not develop nonequilibrium 
features~\cite{NietnerSchallerBrandesPRA2014, GallegoMarcosEtAlPRA2014, SchallerNietnerBrandesNJP2014, 
GrenierGeorgesKollathPRL2014, SekeraBruderBelzigPRA2016, GrenierKollathGeorgesCRP2016}. 

Crucially, we find that using the more adequate second law, Eq.~(\ref{eq Clausius}) instead of 
Eq.~(\ref{eq Clausius 2}), reveals surprising insights: finite-time information erasure and heat engines have higher 
efficiencies than previously thought. These general results follow from the central relation: 
\begin{equation}\label{eq central}
 \Sigma' - \Sigma 
 = D\big(\pi_B[T(\tau)]\big|\pi_B[T(0)]\big) \ge 0.
\end{equation}
Here, $D(\rho|\sigma) \equiv \mbox{tr}\{\rho(\ln\rho-\ln\sigma)\}$ is the always positive quantum relative entropy, 
measuring the statistical distance between two states $\rho$ and $\sigma$. The derivation of Eq.~(\ref{eq central}) 
is simple. Let $S_B(T)$ denote the von Neumann entropy of $\pi_B(T)$. By virtue of 
definition~(\ref{eq def temperature}), which is valid even out of equilibrium, we obtain the relation 
$-\dbar Q(t)/T(t) = dS_B[T(t)]$ and hence $-\int\dbar Q(t)/T(t) = S_B[T(\tau)] - S_B[T(0)]$. Standard manipulations 
then imply Eq.~(\ref{eq central}). The supplemental material (SM) lists all the details of the 
derivation~\footnote{The SM derives Eq.~(\ref{eq central}) in detail, provides information and additional numerics 
for the Landauer erasure protocol, explains the swap engine in detail, demonstrates that our results remain valid 
beyond the steady state regime, and confirms our conclusions if additional information about the bath is available. 
It also contains Ref.~\cite{SafranekEtAlArXiv2020}.}. 

The central result~(\ref{eq central}) tells us that the entropy production $\Sigma$ is smaller than what one would 
naively expect based on $\Sigma'$. Thus, the process is actually \emph{less irreversible} in reality. Physically 
speaking, we can explain Eq.~(\ref{eq central}) by pointing out that the available information about the heat flow $Q$ 
is taken fully into account in $\Sigma$ but only partially in $\Sigma'$. The inequality $\Sigma'\ge0$ reflects 
the second law for an observer who \emph{ignores} that the bath is finite. However, if one already knows the 
heat flow $Q$, one can use it to gain a more accurate description via the definition~(\ref{eq def temperature}) 
of a time-dependent temperature. Thus, $\Sigma$ efficiently uses the available information and the loss 
in predictive power resulting from ignoring the finiteness of the bath is quantified 
by the relative entropy in Eq.~(\ref{eq central}). 
Remarkably, since the effective temperature $T(\tau)$ is in one-to-one correspondence to the bath energy, 
Eq.~(\ref{eq central}) also reveals that the computation of $\Sigma$ does \emph{not} require more information than 
the computation of $\Sigma'$: both are uniquely fixed by knowing $T(0)$ and $Q(\tau)$. 

Another interpretation of Eq.~(\ref{eq central}) is the following. Suppose that we have an \emph{additional} infinitely 
large \emph{superbath} at our disposal with fixed temperature $T(0)$. After the finite bath has interacted 
with the system, it is out of equilibrium with respect to this superbath if $T(\tau) \neq T(0)$. This nonequilibrium 
situation can be used to extract work. The maximum extractable work equals the change in free energy: 
$W_\text{ext}^\text{max} = F_B[T(\tau)] - F_B[T(0)]$~\cite{BinderEtAlBook2018,LandiPaternostroArXiv2020}. Here, 
$F_B(T) \equiv E_B(T) - T(0) S_B(T)$ denotes the nonequilibrium free energy with respect to the reference temperature 
$T(0)$. Note that, even if $\rho_B(t) \neq \pi_B[T(t)]$, $F_B[T(\tau)]$ correctly quantifies the nonequilibrium 
free energy at time $\tau$ based on \emph{our} level of description, which assumes only the bath energy to be 
known (in case of additional information, more work can be extracted). We find 
\begin{equation}\label{eq work extract}
 W_\text{ext}^\text{max} = T(0)(\Sigma' - \Sigma) \ge 0.
\end{equation}
Thus, if we demand that the bath in our description gets reset after each process to its initial temperature, 
Eq.~(\ref{eq central}) tells us that we can always use this reset stage to extract useful work, which remains 
unaccounted for in Eq.~(\ref{eq Clausius 2}). 

In the following, we explicitly demonstrate the use and benefit of Eq.~(\ref{eq Clausius}) for two relevant cases: 
information erasure and heat engines. 

\begin{figure*}%[h]
 \centering\includegraphics[width=0.95\textwidth,clip=true]{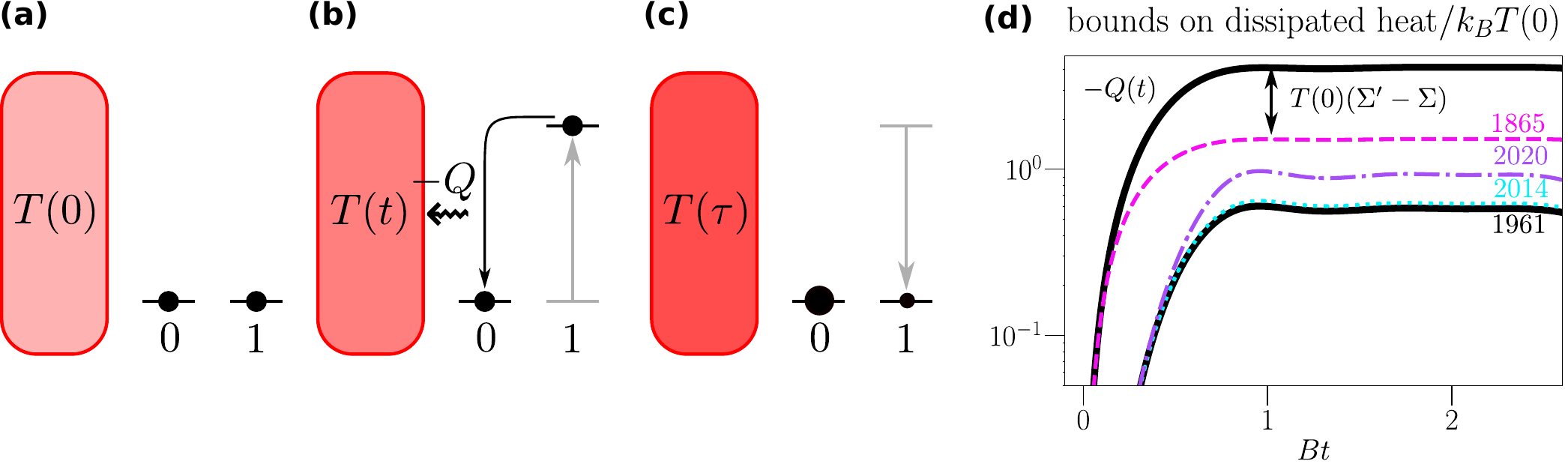}
 \label{fig Landauer} 
 \caption{Landauer erasure protocol. \bb{(a)} A two-level system in a maximally mixed state 
 $\rho_S(0) = (|0\rl0|+|1\rl1|)/2$ with degenerate energies is in contact with a finite bath at initial 
 temperature $T(0)$. \bb{(b)} The energy of state 1 is lifted to increase the probability for a jump $1\rightarrow0$ 
 emitting heat $-Q$ into the bath, thereby increasing its temperature. \bb{(c)} After some time $\tau$, 
 the energy of state 1 is returned to its initial value. The final state has a population imbalance towards the state 
 0 (populations are indicated by the size of the black disks). \bb{(d)} Plot exemplifying different inequalities 
 in the literature for a system-bath Hamiltonian with $H_S = B \sigma_S^z$, 
 $H_B = B\sum_{j=1}^N \sigma^z_j + J \sum_{j=1}^{N-1} (\sigma^x_j\sigma^x_{j+1} + \sigma^y_j\sigma^y_{j+1})$ 
 and $V = J (\sigma^x_S\sigma^x_1 + \sigma^y_S\sigma^y_1)$. We chose $B = J = 1$ ($\hbar\equiv1$) and 
 $T(0) = 1/4$, see the SM for further details and plots. We plot (in logarithmic scale) the total heat 
 $-Q(t)$ (thick black line on top) over dimensionless time $B t$ and compare it with Landauer's bound 
 $-T(0)\Delta S_S(t)$ from 1961~\cite{LandauerIBM1961} (lower thick black line), the bounds from 
 2014~\cite{ReebWolfNJP2014} (dotted turquoise) and 2020~\cite{TimpanaroSantosLandiPRL2020} 
 (dash-dotted purple) as well as the term $-T(0)\int_0^t \dbar Q(s)/T(s)$ labeled by 1865~\cite{Clausius1865} 
 (dashed magenta). The latter is investigated here for the first time and the difference to $-Q(t)$ is given by 
 Eq.~(\ref{eq central}) times $T(0)$. The bound from 2015~\cite{GooldPaternostroModiPRL2015} is always zero for a 
 maximally mixed initial state (see SM). }
\end{figure*}

Erasing one bit of information has become a paradigmatic example for a nonequilibrium thermodynamic process since 
Landauer's groundbreaking work~\cite{LandauerIBM1961}, where he argued that the minimal heat dissipation is 
$-Q(\tau) \ge T(0)\ln 2$ (recall that heat is defined positive if it increases the system energy). More generally, 
Eq.~(\ref{eq Clausius 2}) implies $-Q(\tau) \ge -T(0)\Delta S_S(\tau)$, which coincides with Landauer's principle when 
applied to the erasure of one bit of information (see Fig.~\ref{fig Landauer} for a sketch). 

In the future, the design of energy-efficient computers will become important. Hence, recent effort has been also put 
into obtaining tighter bounds on the heat generation  during information erasure for finite 
baths~\cite{ReebWolfNJP2014, GooldPaternostroModiPRL2015, TimpanaroSantosLandiPRL2020}. The physical nature of these 
bounds is, however, less transparent as they are a consequence of mathematical identities not directly related to the 
second law. 

As explained above, for a finite bath the second law is related to Eq.~(\ref{eq Clausius}). Remarkably, we find that 
\begin{equation}\label{eq Landauer our version}
 -Q(\tau) \ge -T(0)\int_0^\tau\frac{\dbar Q(t)}{T(t)} \ge -T(0)\Delta S_S(\tau),
\end{equation}
where the first inequality follows from Eq.~(\ref{eq central}) and the second from Eq.~(\ref{eq Clausius}). The right 
inequality in Eq.~(\ref{eq Landauer our version}) can be faithfully called `Landauer's principle for a finite bath'. 
It is a logical consequence of applying the second law to memory erasure. For illustration, 
Fig.~\ref{fig Landauer} compares the bounds~(\ref{eq Landauer our version}) as well as the bounds 
from Refs.~\cite{ReebWolfNJP2014, GooldPaternostroModiPRL2015, TimpanaroSantosLandiPRL2020} for the example of a 
spin coupled to a spin chain. In the SM, we provide further numerics, demonstrating that there is no unique relation 
between the bounds. 

We now turn to heat engines and extend our analysis to a system in contact with a finite hot and a finite cold bath. 
The initial system-bath state is generalized to $\rho_S(0)\otimes\pi_C[T_C(0)]\otimes\pi_H[T_H(0)]$, where the 
subscript $C/H$ refers to the cold/hot bath. 

For simplicity, we assume that the engine has operated for a sufficient amount of time 
(or cycles) such that its change in entropy $\Delta S_S$ and internal energy $\Delta U_S$ is negligible compared to 
other terms appearing in the first and second law. This is called the \emph{steady state} regime and it is 
well justified if the system is small in comparison with the baths. However, our conclusions do not depend on this 
assumption as shown in the SM. 

Under the conditions spelled out above, the Clausius inequality can be generalized to~\cite{StrasbergWinterArXiv, 
RieraCampenySanperaStrasbergPRXQ2021} 
\begin{equation}\label{eq EP}
 \Sigma = -\int\frac{\dbar Q_C(t)}{T_C(t)} - \int \frac{\dbar Q_H(t)}{T_H(t)} \ge 0,
\end{equation}
where $\dbar Q_{C/H}(t)$ denotes the infinitesimal heat flow from the baths at time $t$. Moreover, also 
Eq.~(\ref{eq Clausius 2}) can be generalized to~\cite{EspositoLindenbergVandenBroeckNJP2010} 
\begin{equation}\label{eq EP 2}
 \Sigma' = -\frac{Q_C(\tau)}{T_C(0)} - \frac{Q_H(\tau)}{T_H(0)} \ge 0.
\end{equation}
The first law reads $W + Q_C + Q_H = 0$ and our goal is to achieve $W<0$, i.e., we want to \emph{extract} work. 

Note that in presence of two baths the difference~(\ref{eq central}) is given by a sum of relative entropies, 
$\Sigma' - \Sigma = D_C + D_H$ with $D_{C/H} \equiv D(\pi_{C/H}[T_{C/H}(\tau)]|\pi_{C/H}[T_{C/H}(0)])$. 
Hence, $\Sigma' - \Sigma \ge 0$ still holds and it is natural to expect that a heat engine has a higher efficiency 
according to Eq.~(\ref{eq EP}) in comparison with Eq.~(\ref{eq EP 2}). 

However, the correct definition of an efficiency for finite baths is subtle and requires some care. The 
standard choice $-W/Q_H$, which implies the Carnot bound, is adapted to the situation of an engine operating between 
two heat baths with a \emph{fixed} temperature. This is not the case here. Instead, we argue that a meaningful 
efficiency can be properly defined in general as follows. Consider a positive entropy production $\Sigma$ 
split into two contributions labeled $A$ and $B$: $\Sigma = A + B \ge 0$. Now, the universal idea behind defining an 
efficiency is that we want to know how much we have to pay in order to extract something useful. Let the useful 
quantity be $A<0$, which has to be compensated by investing $B>0$ such that $\Sigma \ge 0$. Then, we define 
\begin{equation}
 \eta \equiv \frac{-A}{B} = 1 - \frac{\Sigma}{B} \le 1.
\end{equation}
Importantly, independent of $A$, $B$ and $\Sigma$, this efficiency is always bounded by the same number ($= 1$). 
Note that our universal efficiency definition applies to any engine. This includes, e.g., also 
hybrid engines~\cite{ManzanoEtAlPRR2020}, where a similar (but not identical) efficiency was proposed. 

\begin{figure*}%[h]
 \centering\includegraphics[width=0.95\textwidth,clip=true]{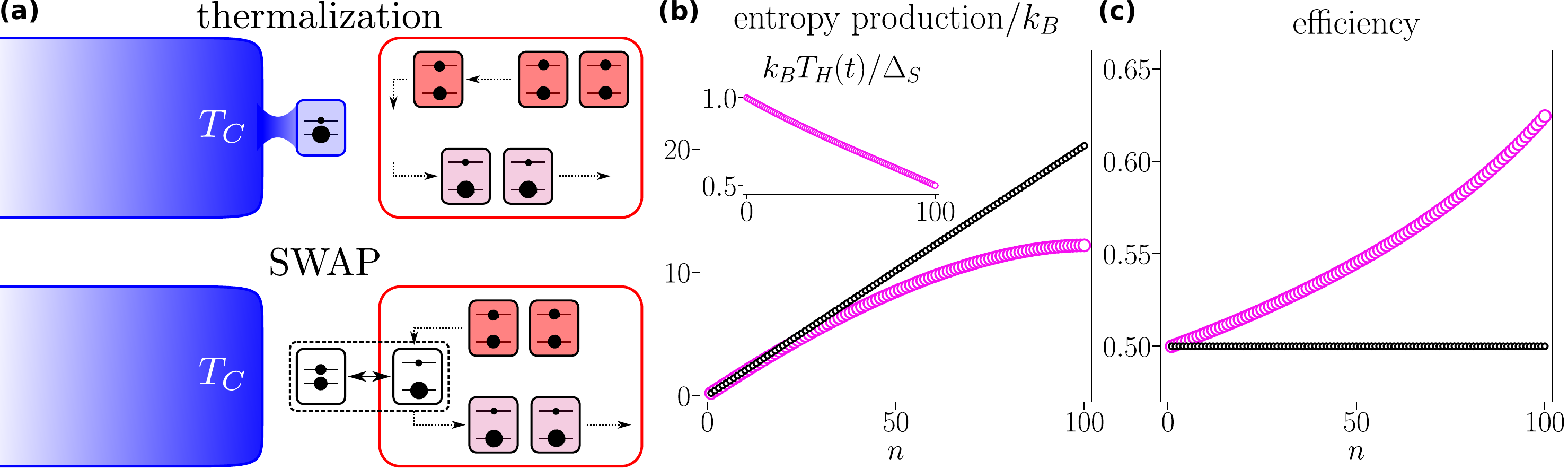}
 \label{fig machine} 
 \caption{\bb{(a)} Sketch of a system (a qubit with energy gap $\Delta_S$) coupled to an infinite cold bath at 
 temperature $T_C$ and a finite hot bath made up of $N$ qubits with energy gap $\Delta_H$. The cycle of the engine 
 consists of two steps. First, a thermalization step with the cold bath such that the system is initialized in 
 the state $\pi_S(T_C)$ (no contact with the hot bath). Second, a short and sudden interaction with a qubit of the 
 hot bath modeled by a swap operation. Each qubit of the hot bath 
 is initially at temperature $T_H(0)$, interacts only \emph{once} with the system, and is afterwards found at a 
 colder temperature (qubit populations are indicated by the size of the black disks). The model is exactly solved in 
 the SM and work extraction is possible if $\Delta_S/\Delta_H > T_C/T_H(0)$. \bb{(b)} We plot $\Sigma$ (large magenta 
 circles) and $\Sigma'$ (small black circles) and the changing temperature $T_H(t)$ of the finite hot bath (inset) as 
 a function of the number of cycles $n$ for $N=100$ qubits in the hot bath. \bb{(c)} Plot of $\eta$ (large magenta 
 circles) and $\eta'$ (small black circles) as a function of $n$. All our plots confirm our conclusions in the main 
 text. Numerical parameters: $\Delta_S = 1$, $\Delta_H = 3/2$, $T_H(0) = 1$, $T_C = 1/3$. }
\end{figure*}

We start by applying this logic to $\Sigma'$. Using the first law, we rewrite Eq.~(\ref{eq EP 2}) as 
\begin{equation}
 \Sigma' = \frac{W(\tau)}{T_C(0)} + \frac{Q_H(\tau)}{T_C(0)} -  \frac{Q_H(\tau)}{T_H(0)} \ge 0.
\end{equation}
Clearly, if operated as a heat engine, the first term is negative and we identify $A' = W(\tau)/T_C(0)$, whereas 
$B' = \eta_C[T_C(0),T_H(0)]Q_H(\tau)/T_C(0)$ is then necessarily positive. Here, we defined the Carnot 
efficiency $\eta_C(T_C,T_H) \equiv 1 - T_C/T_H$. Then, we find 
\begin{equation}
 \eta' = \frac{-A'}{B'} = \frac{1}{\eta_C[T_C(0),T_H(0)]}\frac{-W(\tau)}{Q_H(\tau)} \le 1,
\end{equation}
which simply is a \emph{linearly rescaled} version of the conventional definition. 

Next, we apply this logic to $\Sigma$ and write $\Sigma = A + B$. Now, to make a comparison of efficiencies 
meaningful, we define them with respect to the \emph{same} useful quantity $A = A' = W(\tau)/T_C(0)$. However, the 
$B$-term is now different because the resources invested in order to extract $A$ are differently counted. 
Specifically, we find 
\begin{equation}
 \begin{split}
  B =& \int \frac{\dbar Q_C(t)}{T_C(0)} \eta_C[T_C(0),T_C(t)] \\
  &+ \int \frac{\dbar Q_H(t)}{T_C(0)} \eta_C[T_C(0),T_H(t)],
 \end{split}
\end{equation}
which quantifies a rescaled heat dissipation. In the limit of constant bath temperatures, $T_{C/H}(t) = T_{C/H}(0)$, we 
obtain $B = B'$. More interestingly, using $\eta = -A/B$, our central result~(\ref{eq central}) allows us to conclude 
\begin{equation}
 \eta' = \left(1 - \frac{T_C(0)(D_C+D_H)}{\eta_C[T_C(0),T_H(0)]Q_H(\tau)}\right)\eta \le \eta.
\end{equation}

We recapitulate our logic used to arrive at the general conclusion $\eta \ge \eta'$. We started from two different 
inequalities $\Sigma \ge 0$ and $\Sigma' \ge 0$. In both inequalities, the \emph{same} microsopically defined heat and 
work fluxes enter, but they are bounded in \emph{different} ways. Based on these two inequalities, we constructed two 
efficiencies $\eta$ and $\eta'$. Importantly, (i) these efficiencies are both bounded by 1 and (ii) both quantify 
the amount of resources needed to extract the same quantity $A = W/T_C(0)$. They are therefore comparable and we have 
shown in full generality that $\eta \ge \eta'$ arises as a consequence of $\Sigma' \ge \Sigma$. 
We also add that, instead of fixing the extracted resources $A=A'$ to be the same [condition (ii)], we could likewise 
fix the invested resources $B = B'$. It is easy to check that our conclusion remains: $\eta \ge \eta'$. 

To illustrate the above points, we have numerically simulated a heat engine in contact with an infinite cold bath and 
a finite hot bath. Our setup combines the idea of a swap engine~\cite{CampisiPekolaFazioNJP2015} with the framework of 
repeated interactions~\cite{StrasbergEtAlPRX2017} and is explained in detail in Fig.~\ref{fig machine} and the SM. 
Our numerical observations in Fig.~\ref{fig machine} confirm our general claims. For other studies of heat 
engines in contact with finite baths see Refs.~\cite{BrandnerSaitoSeifertPRX2015, BrandnerSeifertPRE2016, 
BrandnerSaitoPRL2020, IzumidaOkudaPRL2014, WangPRE2016, PozasKerstjensBrownHovhannisyanNJP2018}. 

Before concluding, we emphasize another subtle point. We stressed above that Clausius' inequality is identical 
to the change in thermodynamic entropy of the universe for an \emph{ideal} heat bath, which is well described 
by a macroscopic temperature $T(t)$. However, via definition~(\ref{eq def temperature}), Eq.~(\ref{eq Clausius}) 
remains microscopically valid for any bath state, but in this case $-\dbar Q(t)/T(t)$ no longer strictly 
relates to a change in bath entropy. Remarkably, it is possible to generalize the second law to also account for 
information about the (coarse-grained) \emph{distribution} of bath energies~\cite{StrasbergWinterArXiv, 
RieraCampenySanperaStrasbergPRXQ2021}. For the same reasons as explained below Eq.~(\ref{eq central}), the 
corresponding efficiency with respect to this refined second law is \emph{even higher} than $\eta$, which we show 
explicitly in the SM. Thus, having access to further information in form of the nonequilibrium distribution of 
bath energies offers additional interesting benefits in unison with other recent findings~\cite{HajilooEtAlPRB2020}. 
Note, however, that the present approach relies arguably only on the minimal information required; namely, 
the average bath energy $E_B(\tau)$ or (which is in one-to-one correspondence) its temperature $T(\tau)$. 

To conclude, 155 years ago, Clausius wrote down a remarkable inequality, which remained unappreciated in quantum 
thermodynamics. Here, we emphasized that the original Clausius inequality~(\ref{eq Clausius}) correctly quantifies the 
second law for a much larger class of situations than the conventionally employed inequality~(\ref{eq Clausius 2}). 
We showed that Eq.~(\ref{eq Clausius}) can be fruitfully used to study nanomachines in contact with finite baths and, 
importantly, it is easy to apply in computations as it only relies on the knowledge of the bath energy. 
Whether it is simple to experimentally measure the bath energy depends on the platform. Here, it could be 
helpful to develop thermometry schemes~\cite{MehboudiSanperaCorreaJPA2019}, which are adapted to the nonequilibrium 
temperature defined in Eq.~(\ref{eq def temperature}). Finally and most remarkably, the ancient 
Clausius inequality offers the insight that nanoscale engines are more efficient than previously anticipated. 

%%%%%%%%%%%%%%%%%%%%%%%%%%%%%%%%%%%%%%%%%%%%%%%%%%%%%%%%%%%%%%%%%%%%%%%%%%%%%%%%%%%%%%%%%%%%%%%%%%%%%%%%%%%%%%%%%%%%%%%%
\emph{Acknowledgements.---}We are grateful to Anna Sanpera for many stimulating discussions and valuable comments on 
the manuscript. PS is financially supported by the DFG (project STR 1505/2-1). MGD acknowledges financial support from 
Secretaria d'Universitats i Recerca del Departament d'Empresa i Coneixement de la Generalitat de Catalunya, co-funded 
by the European Union Regional Development Fund within the ERDF Operational Program of Catalunya (project QuantumCat, 
ref. 001-P-001644). ARC acknowledges financial support from the Generalitat de Catalunya: AGAUR FI2018-B01134. 
All authors acknowledge financial support from the Spanish Agencia Estatal de Investigaci\'on, project 
PID2019-107609GB-I00, the Spanish MINECO FIS2016-80681-P (AEI/FEDER, UE), and Generalitat de Catalunya CIRIT 
2017-SGR-1127. 

%%%%%%%%%%%%%%%%%%%%%%%%%%%%%%%%%%%%%%%%%%%%%%%%%%%%%%%%%%%%%%%%%%%%%%%%%%%%%%%%%%%%%%%%%%%%%%%%%%%%%%%%%%%%%%%%%%%%%%%%

\bibliography{/home/philipp/Documents/references/books,/home/philipp/Documents/references/open_systems,/home/philipp/Documents/references/thermo,/home/philipp/Documents/references/info_thermo,/home/philipp/Documents/references/general_QM,/home/philipp/Documents/references/math_phys,/home/philipp/Documents/references/equilibration}
%\bibliography{/home/wiwi/Documents/references/books,/home/wiwi/Documents/references/open_systems,/home/wiwi/Documents/references/thermo,/home/wiwi/Documents/references/info_thermo,/home/wiwi/Documents/references/general_QM,/home/wiwi/Documents/references/math_phys,/home/wiwi/Documents/references/general_refs,/home/wiwi/Documents/references/equilibration}
%\bibliography{books,open_systems,thermo,general_QM,info_thermo,math_phys}

\onecolumngrid
%%%%%%%%%%%%%%%%%%%%%%%%%%%%%%%%%%%%%%%%%%%%%%%%%%%%%%%%%%%%%%%%%%%%%%%%%%%%%%%%%%%%%%%%%%%%%%%%%%%%%%%%%%%%%%%%%%%%%%%%
\appendix

%%%%%%%%%%%%%%%%%%%%%%%%%%%%%%%%%%%%%%%%%%%%%%%%%%%%%%%%%%%%%%%%%%%%%%%%%%%%%%%%%%%%%%%%%%%%%%%%%%%%%%%%%%%%%%%%%%%%%%%%
\section{SUPPLEMENTARY MATERIAL}

We here detail in chronological order additional information concerning (A) the derivation of Eq.~(4) in the maintext, 
(B) the Landauer erasure protocol used for the numerics, (C) the swap engine coupled to a repeated interactions bath, 
(D) the fact that our conclusions remain true even if the system has not yet reached a steady state, (E) improved 
efficiencies if additional information about the nonequilibrium probability distribution of the bath energies is 
available. Here, we sometimes make use of the inverse temperature $\beta(t) = 1/T(t)$ whenever convenient. 

%%%%%%%%%%%%%%%%%%%%%%%%%%%%%%%%%%%%%%%%%%%%%%%%%%%%%%%%%%%%%%%%%%%%%%%%%%%%%%%%%%%%%%%%%%%%%%%%%%%%%%%%%%%%%%%%%%%%%%%%
\section{(A) Details concerning the derivation of Eq.~(4) in the maintext}

We formulate our derivation using the inverse temperature $\beta(t)$ and assume for the moment that it changes 
in a differentiable way. We start by considering the integral 
\begin{equation}
 -\int\frac{\dbar Q(t)}{T(t)} = \int_0^\tau dt \beta(t) \frac{dE_B(t)}{dt},
\end{equation}
where we used $-\dbar Q(t) = dE_B(t) = \mbox{tr}_B\{H_B[\rho_B(t+dt) - \rho_B(t)]\}$. The next step requires to 
confirm that 
\begin{equation}\label{eq heat capacity}
 \frac{dE_B(t)}{dt} = -\dot\beta(t)\mbox{Var}[H_B,\beta(t)], ~~~ \dot\beta(t) \equiv \frac{d\beta(t)}{dt}, ~~~ 
 \mbox{Var}(H_B,\beta) \equiv \mbox{tr}_B\{H_B^2\pi_B(\beta)\} - \mbox{tr}_B\{H_B\pi_B(\beta)\}^2.
\end{equation}
Note that the first equation in Eq.~(\ref{eq heat capacity}) is equivalent to the known result for the heat capacity of 
the canonical ensemble, $C_B \equiv dE(t)/dT(t) = \mbox{Var}[H_B,\beta(t)]/T(t)^2$, after noting that 
$\dot\beta(t) = -\dot T(t)/T(t)^2$. 

Now, let $S_B(\beta) \equiv -\mbox{tr}_B\{\pi_B(\beta)\ln\pi_B(\beta)\}$ denote the von Neumann entropy of a canonical 
equilibrium state at inverse temperature $\beta$. Using 
$d_t S_\text{vN}[\rho(t)] = -\mbox{tr}\{[d_t\rho(t)]\ln\rho(t)\}$ and manipulations similar to those required to 
obtain Eq.~(\ref{eq heat capacity}), we find that 
\begin{equation}
 \frac{dS_B[\beta(t)]}{dt} = -\beta(t)\dot\beta(t)\mbox{Var}[H_B,\beta(t)].
\end{equation}
Taken together, we confirm that 
\begin{equation}\label{eq SM heat entropy}
 \int_0^\tau dt \beta(t) \frac{dE_B(t)}{dt} = \int_0^\tau dt\frac{dS_B[\beta(t)]}{dt} 
 = S_B[\beta(\tau)] - S_B[\beta(0)].
\end{equation}

Next, we note the general identity 
\begin{equation}
 -\frac{Q(\tau)}{T(0)} 
 = \mbox{tr}_B\{\pi_B[\beta(0)]\ln\pi_B[\beta(0)]\} - \mbox{tr}_B\{\pi_B[\beta(\tau)]\ln\pi_B[\beta(0)]\}.
\end{equation}
Combining this with Eq.~(\ref{eq SM heat entropy}), we confirm 
\begin{equation}
 \Sigma' - \Sigma = -\frac{Q(\tau)}{T_0} + \int_0^\tau \frac{\dbar Q(t)}{T(t)} 
 = D\big[\pi_B[\beta(\tau)]\left|\pi_B[\beta(0)]\big]\right.,
\end{equation}
which proves Eq.~(4) in the maintext of the main text. 

Two comments are important. First, we preferred to work with $\beta(t)$ instead of $T(t)$. This choice is 
related to the phenomenon that for a bath with a finite Hilbert space (note that there can be \emph{finite} baths with 
an \emph{infinite} Hilbert space, e.g., a collection of particles in a box) the temperature $T(t)$ becomes negative 
for a population inverted state. Importantly, when the state changes continuously from a state $\rho_B(t_-)$ 
\emph{without} population inversion to a state $\rho_B(t_+)$ \emph{with} population inversion, the associated 
temperature $T(t)$ changes via definition~(3) in the maintext from $T(t_-) = \infty$ to $T(t_+) = -\infty$, 
i.e., there is a sudden and discontinuous jump in the temperature. However, this jump can be avoided when working with 
the inverse temperature, which changes continuously from $\beta(t_-) > 0$ to $\beta(t_+) < 0$. Therefore, by working 
with the inverse temperature, we demonstrated that our conclusions remain true even for exotic negative temperature 
states in the bath. 

Second, we comment on the assumption that $\beta(t)$ needs to be differentiable. In fact, since unitary dynamics 
generated by the Liouville-von Neumann equation are differentiable, there are good reasons to expect that the effective 
(inverse) temperature of the bath defined via Eq.~(3) in the maintext also changes in a differentiable way. 
Importantly, since the relation $dS_B[\beta(t)] = \beta(t) dE_B[\beta(t)]$ has to hold only under the integral, our 
central result~(4) in the maintext remains valid if there is a finite number of times for which $\beta(t)$ is not 
differentiable, but still continuous. We indeed observe this behaviour in Sec.~(B), where $\beta(t)$ shows 
(non-differentiable) spikes without invalidating Eq.~(4) in the maintext. Finally, cases where $\beta(t)$ is not even 
continuous can be only generated by singular cases such as, e.g., a Hamiltonian with a time-dependence described by 
a Dirac delta function. Strictly speaking, these models are, of course, unphysical. Nevertheless, we investigate this 
case in detail in Sec.~(C) for the swap engine. Indeed, the swap engine models the swap operation as happening 
instantaneously, which generates a discontinuous evolution of $\beta(t)$. Despite this feature, we find that the 
central inequality $\Sigma' - \Sigma \ge 0$ continuous to hold and that the difference between $\Sigma' - \Sigma$ and 
$D\{\pi_B[\beta(\tau)]|\pi_B[\beta(0)]\}$ quickly becomes negligibly small. 

%%%%%%%%%%%%%%%%%%%%%%%%%%%%%%%%%%%%%%%%%%%%%%%%%%%%%%%%%%%%%%%%%%%%%%%%%%%%%%%%%%%%%%%%%%%%%%%%%%%%%%%%%%%%%%%%%%%%%%%%
\section{(B) Details of the Landauer erasure protocol}

We first review the bounds of Refs.~\cite{ReebWolfNJP2014, GooldPaternostroModiPRL2015, TimpanaroSantosLandiPRL2020} 
before specifying the details of our numerical studies. 

%%%%%%%%%%%%%%%%%%%%%%%%%%%%%%%%%%%%%%%%%%%%%%%%%%%%%%%%%%%%%%%%%%%%%%%%%%%%%%%%%%%%%%%%%%%%%%%%%%%%%%%%%%%%%%%%%%%%%%%%
\subsection{Bounds to the dissipated heat}

The bound of Ref.~\cite{ReebWolfNJP2014} reads 
\begin{equation}\label{eq Reeb Wolf}
 -Q(\tau) \ge -T(0)\Delta S_S(\tau) + \frac{2T(0)\Delta S_S^2(\tau)}{\ln^2(d-1)+4},
\end{equation}
where $d$ is the dimension of the \emph{bath} Hilbert space. 

The bound of Ref.~\cite{GooldPaternostroModiPRL2015} is based on the fact that the dynamics of the bath can be 
written as $\rho_B(t) = \sum_\alpha A_\alpha(t) \rho_B(0) A_\alpha^\dagger(t)$, where the bath operators $A_\alpha(t)$ 
are commonly called Kraus operators. Defining $\bb A(t) \equiv \sum_\alpha A_\alpha(t) A_\alpha^\dagger(t)$, the bound 
reads 
\begin{equation}\label{eq Goold Paternostro Modi}
 -Q(t) \ge -T(0)\ln\mbox{tr}_B\{\bb A(t)\rho_B(0)\}.
\end{equation}
We now demonstrate that this bound is zero for the case of Landauer erasure, i.e., whenever the system starts in a 
maximally mixed state. To this end, we note that the operators $A_\alpha(t)$ are microscopically defined as 
$A_\alpha(t) = \sqrt{\lambda_j}\langle s_k|U_{SB}(t)|s_j\rangle$, where $U_{SB}(t)$ is the global system bath unitary 
and the initial state of the system was decomposed as $\rho_S(0) = \sum_j \lambda_j |s_j\rl s_j|$. Thus, we see that 
the index $\alpha$ is actually a double-index $\alpha = (j,k)$. Now, if $\rho_S(0)$ is a maximally mixed state, then 
$\lambda_j = 1/\sqrt{d_S}$, where $d_S$ is the dimension of the system Hilbert space. Consequently, 
\begin{equation}
 \bb A(t) = \sum_\alpha A_\alpha(t) A_\alpha^\dagger(t) 
 = \sum_{j,k} \frac{1}{d_S}\langle s_k|U_{SB}(t)|s_j\rangle \langle s_j|U_{SB}^\dagger(t)|s_k\rangle 
 = \sum_k \frac{1}{d_S}\langle s_k|U_{SB}(t)U_{SB}^\dagger(t)|s_k\rangle
 = \sum_k \frac{1}{d_S} = 1,
\end{equation}
where $1$ denotes the identity operator in the bath space. Thus, $-\ln\mbox{tr}_B\{\bb A(t)\rho_B(0)\} = -\ln(1) = 0$. 

Next, we turn to the bound from Ref.~\cite{TimpanaroSantosLandiPRL2020}. Interestingly, this bound also employs the 
temperature definition~(3) in the maintext. We introduce the two functions 
$f[\beta(t)] \equiv \mbox{tr}_B\{H_B\pi_B[\beta(t)]\} - \mbox{tr}_B\{H_B\pi_B[\beta(0)]\}$ and 
$g[\beta(t)] \equiv S_B[\beta(t)] - S_B[\beta(0)]$, which determine the energy and entropy change with 
respect to the fictitious equilibrium state of the bath at inverse temperature $\beta(t)$ starting from $\beta(0)$. 
It was then found that 
\begin{equation}\label{eq Landi bound}
 -Q(t) \ge (f\circ g^{-1})[-\Delta S_S(t)].
\end{equation}
Here, $(f\circ g^{-1})(x) \equiv f[g^{-1}(x)]$ denotes the concatenation of two functions. 

%%%%%%%%%%%%%%%%%%%%%%%%%%%%%%%%%%%%%%%%%%%%%%%%%%%%%%%%%%%%%%%%%%%%%%%%%%%%%%%%%%%%%%%%%%%%%%%%%%%%%%%%%%%%%%%%%%%%%%%%
\subsection{Model 1: Spin coupled to a spin chain}

The total system-bath Hamiltonian $H_S + H_B + V$ is specified by 
\begin{equation}
 H_S = B_0 \sigma_z^S, ~~~ H_B = B\sum_{j=1}^N \sigma_z^{(j)} 
 + J \sum_{j=1}^{N-1} (\sigma_x^{(j)}\sigma_x^{(j+1)} + \sigma_y^{(j)}\sigma_y^{(j+1)}), ~~~ 
 V = J_0 (\sigma_x^{S}\sigma_x^{(1)} + \sigma_y^{S}\sigma_y^{(1)}).
\end{equation}
Here, $\sigma_{x,y,z}$ denotes the usual Pauli matrix, $N$ the number of spins in the bath and 
$B_0, B, J_0$ and $J$ are real-valued parameters. The time-evolution of the system-bath state is implemented 
by exact numerical integration. 

\begin{figure*}%[b]
 \centering\includegraphics[width=0.85\textwidth,clip=true]{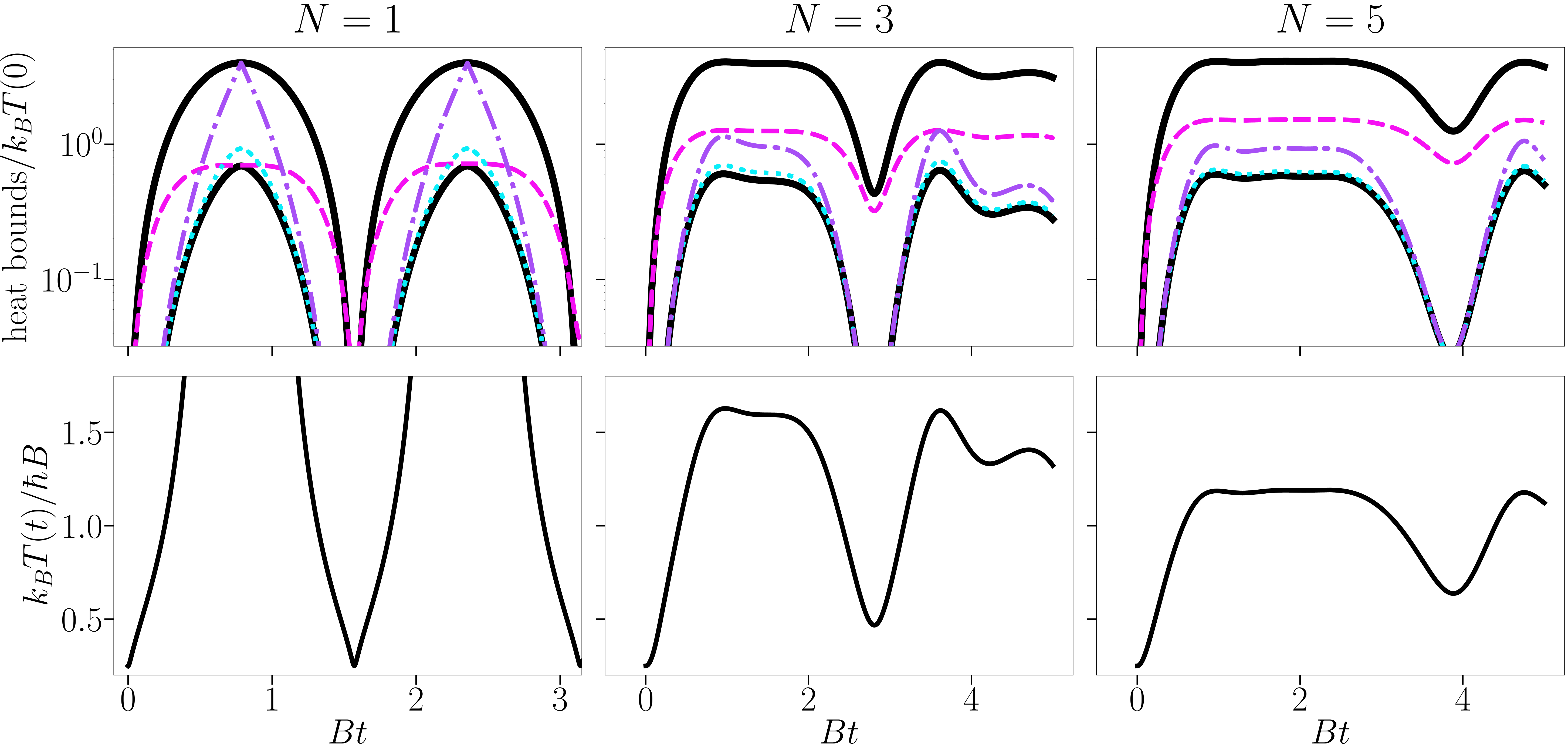}
 \label{fig spin 1 SM} 
 \caption{\bb{First row:} As a function of dimensionless time $B t$ for $N=1$ (left), $N=3$ (center) and $N=5$ 
 (right) spins in the bath, plot of the dissipated heat $-Q(t)$ and Landauer's original lower bound 
 $-T(0)\Delta S_S(t)$ (thick solid black lines) and the bound~(\ref{eq Reeb Wolf})~\cite{ReebWolfNJP2014} (dotted 
 turquoise line), the bound~(\ref{eq Landi bound})~\cite{TimpanaroSantosLandiPRL2020} (dash-dotted purple line) and 
 our bound~(6) in the maintext (dashed magenta line). Note that we used a logarithmic scale for better 
 visibility. \bb{Second row:} Again as a function of time for $N=1$ (left), $N=3$ (center) and $N=5$ (right) spins 
 in the bath, plot of the time-evolution of the (dimensionless) nonequilibrium temperature $T(t)$. 
 Numerical parameters: $B = B_0 = 1$, $J = J_0 = 1$, $T(0) = 1/4$. }
\end{figure*}

We start by giving additional numerical results concering the Landauer erasure protocol, which starts with a maximally 
mixed state $\rho_S(0) = (|0\rl 0|+|1\rl 1|)/2$, where $|0\rangle$ ($|1\rangle$) denotes the ground (excited) state of 
the system. Figure~\ref{fig spin 1 SM} shows the bounds from Refs.~\cite{LandauerIBM1961, ReebWolfNJP2014, 
TimpanaroSantosLandiPRL2020} and our bound~(6) in the maintext on the total heat dissipation for 
$N = 1,3,5$ spins in the bath. We observe that there is no particular order between our bound and the bounds from 
Refs.~\cite{ReebWolfNJP2014, TimpanaroSantosLandiPRL2020}. Furthermore, for $N=1$ we observe non-differentiable 
`spikes' in the evolution of the temperature $T(t)$. Nevertheless, in unison with our last comment in Sec.~(A), we 
find perfect agreement with our central result~(4) in the maintext (not shown explicitly). 

\begin{figure*}%[h]
 \centering\includegraphics[width=0.85\textwidth,clip=true]{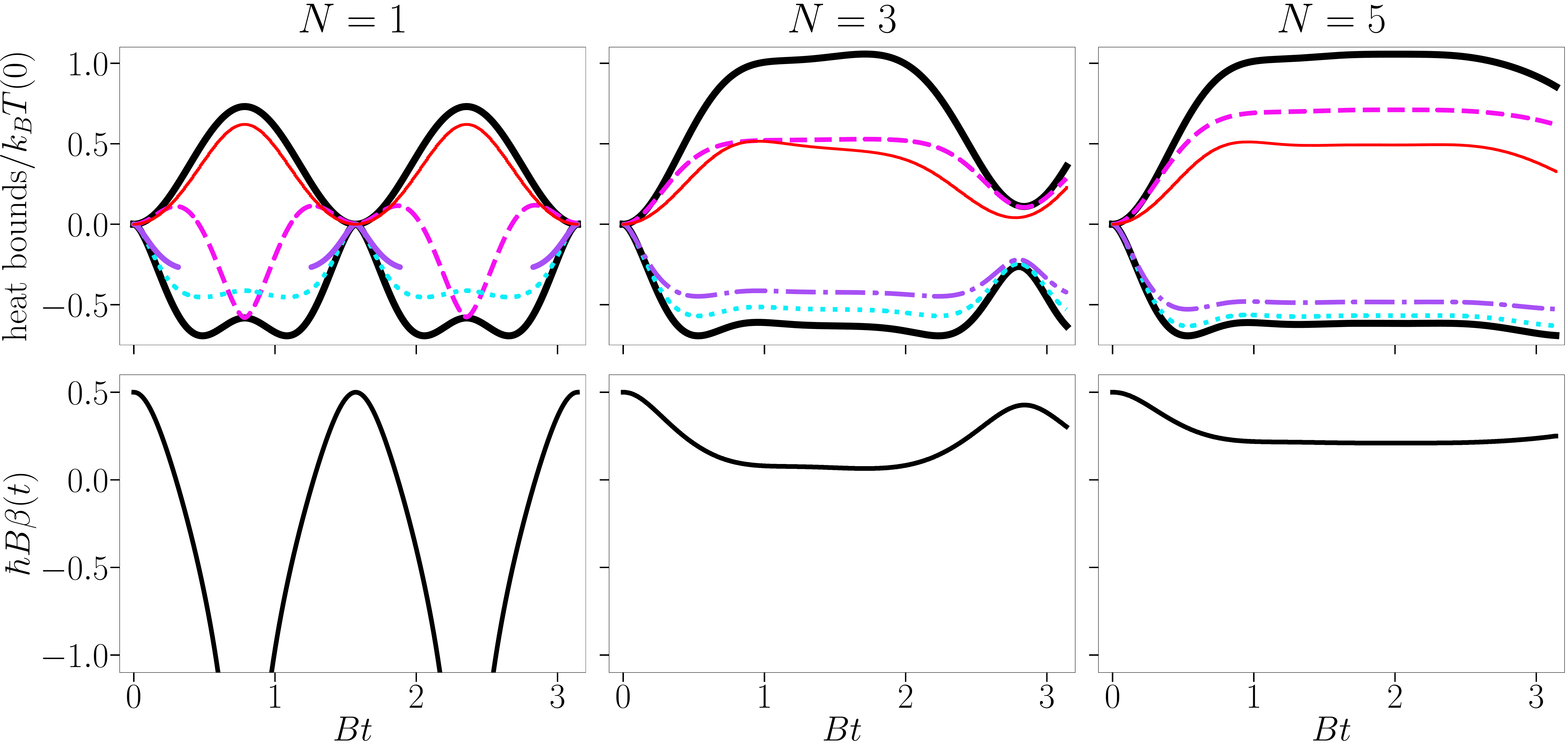}
 \label{fig spin 1 SM 2} 
 \caption{As a function of dimensionless time $B t$ for $N=1$ (left), $N=3$ (center) and $N=5$ (right) spins 
 in the bath, plot of the dissipated heat $-Q(t)$ and Landauer's original lower bound $-T(0)\Delta S_S(t)$ (thick 
 solid black lines) and the bound~(\ref{eq Reeb Wolf})~\cite{ReebWolfNJP2014} (dotted turquoise line), the bound~(\ref{eq Goold Paternostro Modi})~\cite{GooldPaternostroModiPRL2015} (thin red solid line), the 
 bound~(\ref{eq Landi bound})~\cite{TimpanaroSantosLandiPRL2020} (dash-dotted purple line, except for $N=1$ where the 
 bound only exists for a subset of times and where we used a thick purple line) and our bound~(6) in the maintext 
 (dashed magenta line). Parameters are as in Fig.~\ref{fig spin 1 SM} except of a different temperature, $T(0) = 2$, 
 and a different initial system state (an excited state). }
\end{figure*}

Next, we provide additional numerical results when the system starts in a pure state, which we take to be the excited 
state. In this case, we expect the bound from Ref.~\cite{GooldPaternostroModiPRL2015} to be useful and, indeed, 
Fig.~\ref{fig spin 1 SM 2} demonstrates this. Furthermore, Fig.~\ref{fig spin 1 SM 2} illustrates two additional 
insights. First, for $N=1$ the bound of Ref.~\cite{TimpanaroSantosLandiPRL2020} \emph{does not exist} for all times 
since the image of the function $g$ in Eq.~(\ref{eq Landi bound}) is not guaranteed to contain the value 
$-\Delta S_S(t)$. Second, instead of plotting the temperature of the bath in the second row, we plot its \emph{inverse} 
temperature. In fact, for $N=1$ the temperature switches from a positive to a negative value during the evolution, 
which would result in discontinuous jumps in $T(t)$ from $\infty$ to $-\infty$, whereas $\beta(t)$ behaves smoothly 
changes from $0^+$ to $0^-$. 

We remark that we have not performed an exhaustive numerical study, but for all the parameters we checked we found 
numerically the same qualitative behaviour. This is also confirmed by our next model. 

%%%%%%%%%%%%%%%%%%%%%%%%%%%%%%%%%%%%%%%%%%%%%%%%%%%%%%%%%%%%%%%%%%%%%%%%%%%%%%%%%%%%%%%%%%%%%%%%%%%%%%%%%%%%%%%%%%%%%%%%
\subsection{Model 2: Spin coupled to a random matrix environment}

\begin{figure*}%[h]
 \centering\includegraphics[width=0.85\textwidth,clip=true]{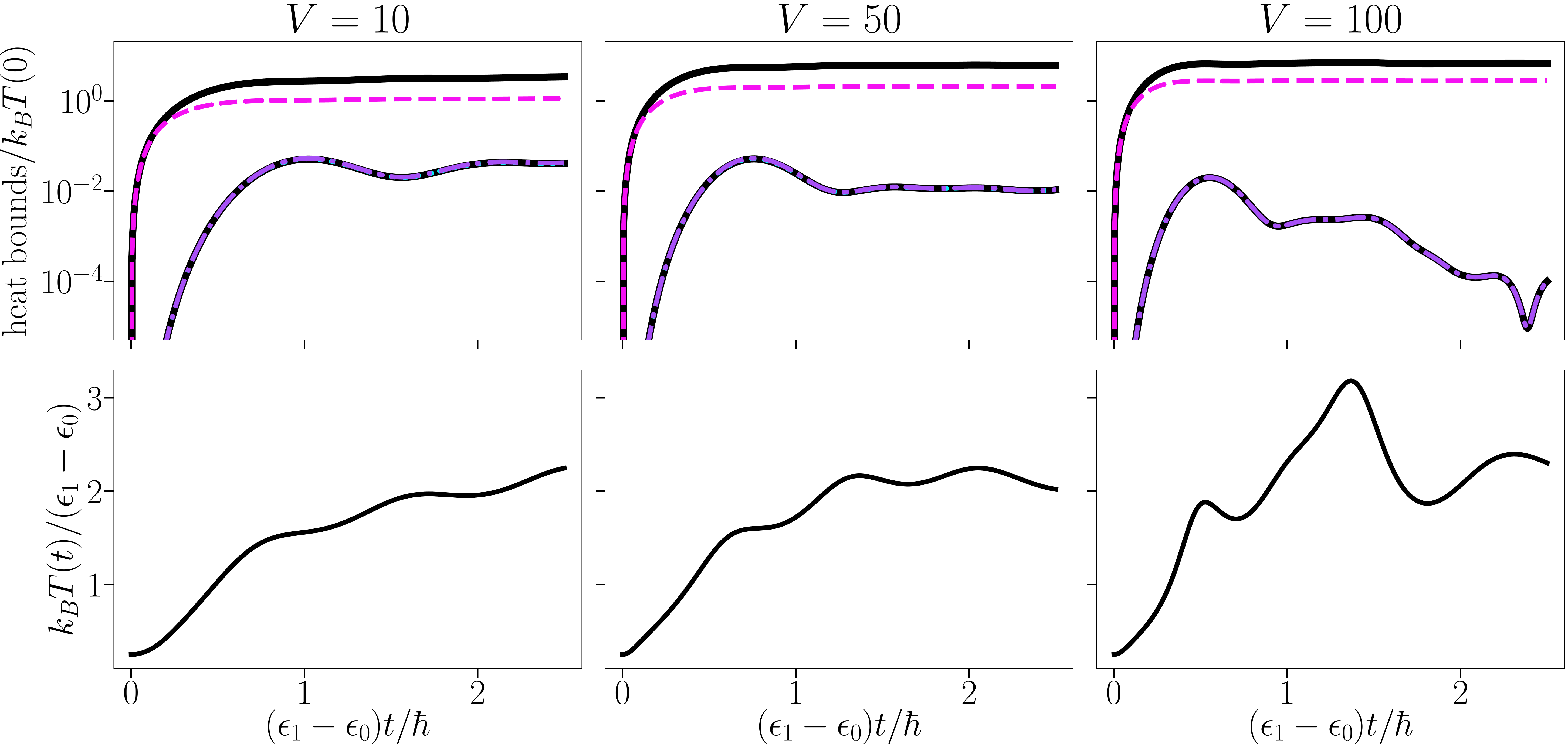}
 \label{fig RM SM} 
 \caption{\bb{First row:} As a function of dimensionless time $(\epsilon_1-\epsilon_0)t/\hbar$ for $V=10$ (left), 
 $V=50$) (center) and $V=100$ (right) levels in the upper band of the bath, plot of the dissipated heat $-Q(t)$ and 
 Landauer's original lower bound $-T(0)\Delta S_S(t)$ (thick solid black lines) and the 
 bound~(\ref{eq Reeb Wolf})~\cite{ReebWolfNJP2014} (dotted turquoise line), the 
 bound~(\ref{eq Landi bound})~\cite{TimpanaroSantosLandiPRL2020} (dash-dotted purple line) and our 
 bound~(6) in the maintext (dashed magenta line). Note that we used a logarithmic scale for better 
 visibility. \bb{Second row:} Again as a function of dimensionless time for $V=10$ (left), $V=50$) (center) and 
 $V=100$ (right) levels in the upper band of the bath, plot of the time-evolution of the (dimensionless) 
 nonequilibrium temperature 
 $T(t)$. Numerical parameters: $\epsilon_1-\epsilon_0 = 1$, $\delta = 1$, $\lambda = 0.3$, $T_0 = 1/4$. }
\end{figure*}

This model was investigated in detail in Sec.~IV of Ref.~\cite{RieraCampenySanperaStrasbergPRXQ2021}. It also 
consists of a spin with Hamiltonian $H_S = \epsilon_0 |0\rl0| + \epsilon_1|1\rl1|$ in contact with a bath with 
Hamiltonian $H_B = \sum_j E_j|j\rl j|$. The bath Hamiltonian is split into two energy \emph{bands} $\C E_0$ and 
$\C E_1$. Each band $\C E_i$ is of width $\delta$ centered around $\epsilon_i$ and contains a number of $V_i$ 
eigenenergies, which are randomly sampled from the interval $[\epsilon_i-\delta/2,\epsilon_i+\delta/2)$. 
Furthermore, we assume $\epsilon_1-\epsilon_0 \ge \delta$ such that the two bands do not overlap. For the numerics, 
we set $V_0 = 10$ and leave $V \equiv V_1$ as a free parameter. Finally, the interaction Hamiltonian is 
\begin{equation}
 V = \lambda \sigma_x \otimes B, ~~~ 
 B = \sum_{E_j\in \C E_0} \sum_{E_k\in \C E_1} c(E_j,E_k) |E_j\rl E_k| + \text{h.c.}
\end{equation}
Here, $c(E_j,E_k)$ is a matrix of independent and identically distributed complex random numbers with zero mean and 
variance $a^2$. The dynamics are generated by formally integrating the exact Schr\"odinger dynamics for \emph{one} 
realization of the coupling matrix $c(E_j,E_k)$, but we observe that the qualitative features of the dynamics do not 
change for different realizations. Despite the bath dimension $d = V_0 + V_1 = 10 + V$ is relatively 
small, the randomness in the coupling helps to mimic a more realistic (i.e., large) heat bath. 
Numerical results are shown in Fig.~\ref{fig RM SM} for the case of Landauer erasure. 

%%%%%%%%%%%%%%%%%%%%%%%%%%%%%%%%%%%%%%%%%%%%%%%%%%%%%%%%%%%%%%%%%%%%%%%%%%%%%%%%%%%%%%%%%%%%%%%%%%%%%%%%%%%%%%%%%%%%%%%%
\section{(C) The swap engine with a repeated interactions bath}

We consider a two-level system with Hamiltonian $H_S = \Delta_S|1\rl1|$, where $|1\rangle$ denotes the excited state 
(and $|0\rangle$ the ground state) as our `working medium'. 

The cold bath is assumed to be an ideal weakly coupled bath, which simply prepares the system in a canonical 
ensemble at temperature $T_C$ in each cycle. We denote this state as $\pi_S(T_C) \equiv e^{-H_S/T_C}/\C Z_S(t_C)$. 

The hot bath is made up of $N$ non-interacting qubits. Each qubit is described by the Hamiltonian 
$H_H = \Delta_H|1\rl1|$ and initialized in the state $\pi_H[T_H(0)] = e^{-H_H/T_H(0)}/\C Z_H[T_H(0)]$. At regular 
time-intervals $k\tau_\text{cycle}$, the $k$'th qubit of the hot bath interacts with the system. This interaction is 
assumed to be fast enough such that the effect of the cold bath can be neglected to lowest order. Furthermore, we 
assume the time $\tau_\text{cycle}$ between two interaction to be large enough such that the system had enough time to 
relax to the thermal state $\pi_S(T_C)$ in between two interactions. Finally, we assume that the interaction between 
the $k$'th bath qubit and the system implements a swap operation described via the unitary operator 
$U_\text{swap}|k,l\rangle = |l,k\rangle$. Here, $|k,l\rangle = |k\rangle_S\otimes|l\rangle_H$ denotes the system qubit 
in state $k$ and the bath qubit in state $l$ ($k,l\in\{0,1\}$). 

We are now in a position, where we can analyze an arbitrary cycle from a thermodynamic perspective. We start with the 
swap operation. The total work equals the change in internal energy of the system and the bath qubit: 
\begin{equation}\label{eq work swap engine}
 W = \mbox{tr}_{SB}\{(H_S+H_B)[U_\text{swap}\pi_S(T_C)\otimes\pi_H[T_H(0)]U_\text{swap}^\dagger - \pi_S(T_C)\otimes\pi_H[T_H(0)]]\}.
\end{equation}
Since there is no work performed on the system during the thermalization processes, we want that this work is 
negative. A quick calculation reveals the explicit expression 
\begin{equation}
 W = 
 (\Delta_S - \Delta_H)\left[\frac{e^{-\Delta_H/T_H(0)}}{\C Z_H[T(0)]} - \frac{e^{-\Delta_S/T_C}}{\C Z_S(T_C)}\right] 
 = (\Delta_S - \Delta_H) \frac{e^{\Delta_S/T_C} - e^{\Delta_H/T_H(0)}}{(e^{\Delta_H/T_H(0)}+1)(e^{\Delta_S/T_C}+1)}.
\end{equation}
The condition for work extraction, $W<0$, follows as 
\begin{equation}\label{eq work extraction condition}
 \frac{T_H(0)}{T_C} > \frac{\Delta_H}{\Delta_S} > 1.
\end{equation}
From Eq.~(\ref{eq work swap engine}) we can also easily calculate the change in system and bath energy during the 
swap operation, which becomes 
\begin{equation}
 \Delta E_S = \Delta_S\left[\frac{e^{-\Delta_H/T_H(0)}}{\C Z_H[T(0)]} - \frac{e^{-\Delta_S/T_C}}{\C Z_S(T_C)}\right], 
 ~~~ \Delta E_H = -\frac{\Delta_H}{\Delta_S}\Delta E_S = -Q_H.
\end{equation}
Here, we equated the change in bath energy with minus the heat flow into it in accordance with our framework above 
and we observe that $W = \Delta E_S - Q_H$, which is the first law during the swap operation. 

During the subsequent equilibration step, the system relaxes back to its initial state by dissipating an amount of 
heat $Q_C = -\Delta E_S$ into the cold bath. This concludes the cycle. Note that the system entropy and energy does 
not change over an \emph{entire} cycle, a property which we also used in the main text. 

Now, the total amount of heat flown and the work after $n$ interactions simply become 
\begin{equation}
 Q_H^\text{tot} = nQ_H, ~~~ Q_C^\text{tot} = nQ_C, ~~~ W^\text{tot} = nW = - Q_C^\text{tot} - Q_H^\text{tot}.
\end{equation}
Furthermore, the two different notions of entropy production become after $n$ interactions
\begin{equation}
 \Sigma = -\frac{nQ_C}{T_C} - \sum_{k=1}^n \frac{Q_H}{T_H(k\tau)} \ge 0, ~~~
 \Sigma' = -\frac{nQ_C}{T_C} -\frac{nQ_H}{T_H(0)} \ge 0,
\end{equation}
which is plotted in Fig.~2(b). Furthermore, the time-dependent temperature of the hot bath 
$T_H(k\tau)$, which is plotted in the inset of Fig.~2(b), can be obtained by solving 
\begin{equation}
 k\frac{\Delta_H}{e^{\Delta_S/T_C}+1} + (N-k)\frac{\Delta_H}{e^{\Delta_H/T_H(0)}+1} = 
 N\frac{\Delta_H}{e^{\Delta_H/T_H(k\tau)}+1},
\end{equation}
where $N$ denotes the total number of qubits in the hot bath. Note that for $k=N$ the temperature of the finite bath 
is exactly $T_H(N\tau) = T_C\Delta_H/\Delta_S$. According to Eq.~(\ref{eq work extraction condition}), this means that 
we can no longer extract work from the system after we have used up all $N$ qubits. The plot of the efficiencies $\eta$ 
and $\eta'$, Fig.~2(c), immediately follows from the above considerations. 

Finally, we return to the second comment made at the end of Sec.~(A) of the SM. There, we found that our central 
result~(4) in the maintext can break down if the bath temperature does not change in a differentiable way, which is 
the case here. Clearly, this behaviour is the more pronounced, the smaller the number of qubits in the cold bath. 
In the extreme case of a bath with a single qubit, the temperature would instantaneously jump from $T_H(0)$ to 
$T_H(\tau) = T_C\Delta_H/\Delta_S < T_H(0)$. Once more, we repeat that this behaviour is \emph{unphysical}. In reality, 
the implementation of the swap operation takes a finite time and then $T_H(t)$ would change continuously. 

\begin{figure*}%[h]
 \centering\includegraphics[width=0.70\textwidth,clip=true]{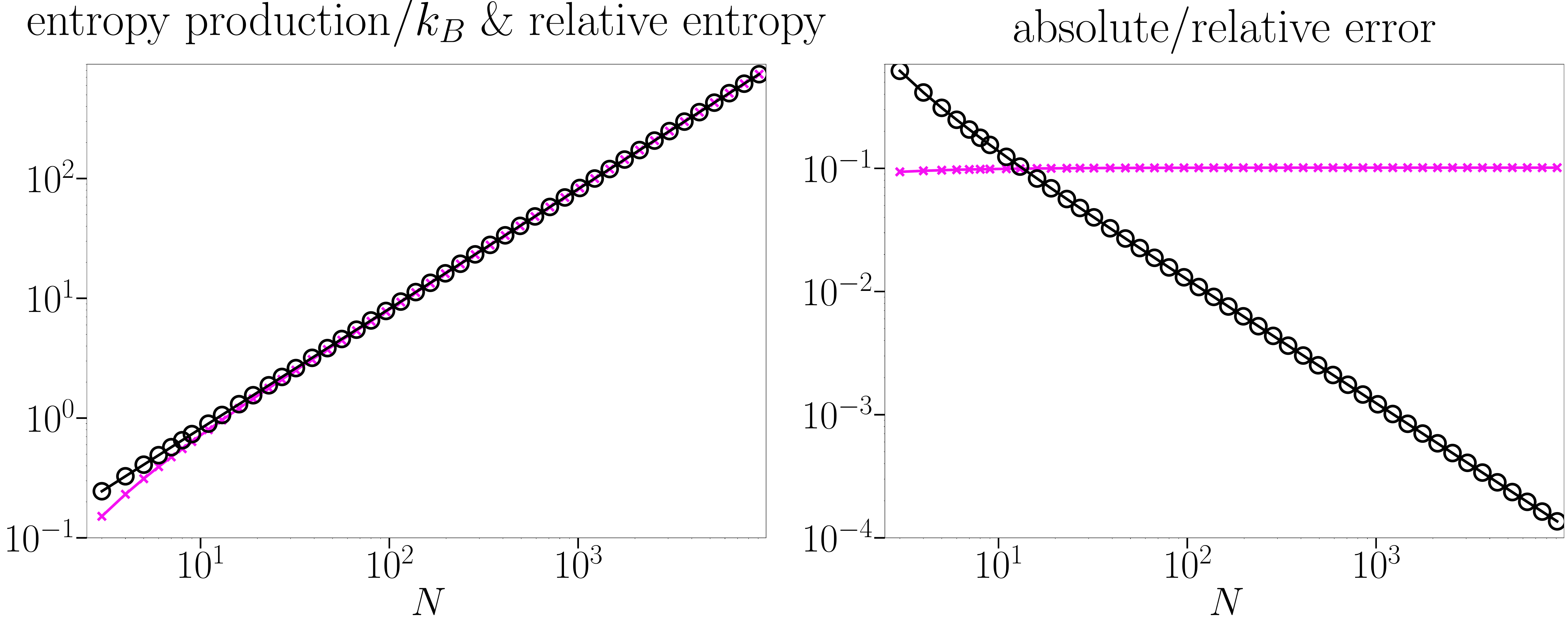}
 \label{fig machine SM} 
 \caption{\bb{Left:} Plot of $\Sigma'(N) - \Sigma(N)$ (pink crosses) and $ND[\pi_B[T_H(N\tau)]|\pi_B[T_H(0)]]$ 
 (black circles) as a function of $N$. \bb{Right:} Plot of $e_\text{abs}(N)$ (pink crosses) and $e_\text{rel}(N)$ 
 (black circles) as a function of $N$. Note that in both plots results are shown on a double-logarithmic scale. 
 Numerical parameters are as for Fig.~2 in the main text: 
 $\Delta_S = 1$, $\Delta_H = 3/2$, $T_H(0) = 1$, $T_C = 1/3$.}
\end{figure*}

Out of curiosity, we ask, however, what happens to our central result~(4) in the maintext for the idealized swap engine 
considered here. The answer is shown in Fig.~\ref{fig machine SM}, where we plot the following quantities. First, 
we plot $\Sigma'(N) - \Sigma(N)$ as a function of the number of qubits $N$ in the hot bath after $N$ cycles have been 
completed (i.e., after each bath qubit has interacted with the system). As one can see (left plot in 
Fig.~\ref{fig machine SM}), the inequality $\Sigma'(N) - \Sigma(N) \ge 0$ continuous to hold. Next, in the same plot 
we also show the relative entropy between the initial state of the hot bath at temperature $T_H(0)$ and the associated 
final equilibrium state at $T_H(\tau)$. Since there are no correlations between the qubits in the hot bath, this 
quantity becomes $N D\{\pi_B[T_H(\tau)]|\pi_B[T_H(0)]\}$, where $\pi_B(T)$ is the canonical ensemble at temperature 
$T$ of a \emph{single} qubit. We observe that, in unison with our central result~(4) in the maintext, the disagreement 
with $\Sigma'(N) - \Sigma(N)$ is very small already for moderate $N$. To quantify this difference more precisely, we 
also display the absolute and relative error in the right plot of Fig.~\ref{fig machine SM}: 
\begin{align}
 e_\text{abs}(N) &\equiv \Sigma'(N) - \Sigma(N) - ND[\pi_B[T_H(N\tau)]|\pi_B[T_H(0)]], \\
 e_\text{rel}(N) &\equiv \frac{\Sigma'(N) - \Sigma(N) - ND[\pi_B[T_H(N\tau)]|\pi_B[T_H(0)]]}{\Sigma'(N) - \Sigma(N)}.
\end{align}
While the absolute error stays constant as a function of $N$,  the \emph{relative error} decreases as $1/N$ because 
the `amount of discontinuity' in $T_H(t)$ gets smaller with increasing $N$. This result is intuitive and reassuring. 

%%%%%%%%%%%%%%%%%%%%%%%%%%%%%%%%%%%%%%%%%%%%%%%%%%%%%%%%%%%%%%%%%%%%%%%%%%%%%%%%%%%%%%%%%%%%%%%%%%%%%%%%%%%%%%%%%%%%%%%%
\section{(D) Nanoscopic heat engines beyond the steady state regime}

If the system has not yet reached a steady state, Eqs.~(7) in the maintext and~(8) in the maintext of the main text generalize 
to 
\begin{align}
 \Sigma &= \Delta S_S(\tau) -\int\frac{\dbar Q_C(t)}{T_C(t)} - \int \frac{\dbar Q_H(t)}{T_H(t)} \ge 0, \\
 \Sigma' &= \Delta S_S(\tau) -\frac{Q_C(\tau)}{T_C(0)} - \frac{Q_H(\tau)}{T_H(0)} \ge 0.
\end{align}
Using the first law $\Delta U_S = Q_C + Q_H + W$, we can rewrite both expressions as 
\begin{equation}
 \Sigma' 
 = \frac{W(\tau)}{T_C(0)} - \frac{\Delta F_S(\tau)}{T_C(0)} + \frac{Q_H(\tau)}{T_C(0)} - \frac{Q_H(\tau)}{T_H(0)} \ge 0.
\end{equation}
Here, we have introduced the change in nonequilibrium free energy 
$\Delta F_S(\tau) = \Delta U_S(\tau) - T_C(0)\Delta S_S(\tau)$ with respect to the reference temperature $T_C(0)$ 
of the initial cold bath. Now, there are different possibilities to split $\Sigma'$ into an $A$- and $B$-term 
depending on which resources we want to convert to each other. However, independent of how we split 
$\Sigma' = A + B$, we can always write 
\begin{equation}
 \Sigma = A + B + \Sigma - \Sigma'.
\end{equation}
Therefore, we obtain the efficiencies 
\begin{equation}
 \eta' = \frac{-A}{B}, ~~~ \eta = \frac{-A}{B + \Sigma - \Sigma'}.
\end{equation}
Since $0 < B + \Sigma-\Sigma' \le B$, we obtain in general the conclusion that $\eta'/\eta \le 1$. Thus, for 
\emph{any process} (not just heat engines) the true efficiency according to the second law~(1) in the maintext 
is larger than the efficiency inferred from Eq.~(2) in the maintext, which was previously asserted to be the 
second law~\cite{BinderEtAlBook2018, LandiPaternostroArXiv2020}. 

%%%%%%%%%%%%%%%%%%%%%%%%%%%%%%%%%%%%%%%%%%%%%%%%%%%%%%%%%%%%%%%%%%%%%%%%%%%%%%%%%%%%%%%%%%%%%%%%%%%%%%%%%%%%%%%%%%%%%%%%
\section{(E) Efficiencies and second law for far-from-equilibrium baths}

In general, a finite bath might be driven out of equilibrium during a thermodynamic process such that it is no longer 
well described by a time-dependent temperature (although this does not seem to be the case in most current 
experiments~\cite{TrotzkyEtAlNP2012, BrantutEtAlScience2012, GringEtAlScience2012, BrantutEtAlScience2013, 
ClosEtAlPRL2016, KaufmanEtAlScience2016, KarimiEtAlNC2020}). In order to describe the bath more accurately, we then 
need, however, more information. Here, we assume this information to be available in terms of the probability 
distribution of the bath energies $p(e_B,t)$ at time $t$ (we only consider one bath here, but the argument generalizes 
to multiple baths). Here, $e_B$ does not necessarily refer to an eigenenergy of the bath Hamiltonian. Instead, 
$p(e_B,t)$ can describe some \emph{coarse-grained} probability distribution as long as the error is small enough 
such that, for instance, the average bath energy is well approximated by 
\begin{equation}
 \mbox{tr}_B\{H_B\rho_B(t)\} \approx \sum_{e_B} e_B p(e_B,t).
\end{equation}
Since $e_B$ does not necessarily refer to a single energy eigenstate, we denote by $V(e_B)$ all microstates which 
give rise to measurement outcome $e_B$. Note, if $p(e_B,t) = \delta_{e_B,e'_B}$ for some $e'_B$, then the state of 
the bath equals the conventional microcanonical ensemble. 

We now define the following notion of entropy, which is known as observational entropy (for a recent review of this 
concept, which goes back to Boltzmann, Gibbs, von Neumann and Wigner, see Ref.~\cite{SafranekEtAlArXiv2020}): 
\begin{equation}
 S_\text{obs}^{E_B}(t) \equiv \sum_{e_B} p(e_B,t)[-\ln p(e_B,t) + \ln V(e_B)].
\end{equation}
As demonstrated in Refs.~\cite{StrasbergWinterArXiv, RieraCampenySanperaStrasbergPRXQ2021}, the second law can then 
be generalized to 
\begin{equation}
 \tilde\Sigma = \Delta S_S(\tau) + S_\text{obs}^{E_B}(\tau) - S_B[T(\tau)] - \int \frac{\dbar Q(t)}{T(t)} \ge 0.
\end{equation}
Since the Gibbs state maximizes entropy for a fixed energy, we infer that 
$S_\text{obs}^{E_B}(\tau) - S_B[T(\tau)]  < 0$. Hence, we obtain the central result 
\begin{equation}
 0 \le \tilde\Sigma \le \Sigma \le \Sigma'.
\end{equation}

Thus, $\tilde\Sigma$ gives an even tighter bound than the conventional Clausius inequality $\Sigma \ge 0$ because 
it takes more information into account. Since the above framework can be generalized to multiple baths (simply by 
adding the contribution of each bath), it is clear that any resulting definition of efficiency $\tilde\eta$ based 
on $\tilde\Sigma$ will satisfy 
\begin{equation}
 1\ge \tilde\eta \ge \eta \ge \eta'.
\end{equation}
Thus, if we assume additional information and control about nonequilibrium features of the bath, efficiencies can be 
even higher. 

Of course, additional information about nonequilibrium features could be also available in other forms. For instance, 
if we additionally know (parts of) the system-bath correlation quantified by the mutual information, an even tighter 
second law emerges~\cite{StrasbergWinterArXiv, RieraCampenySanperaStrasbergPRXQ2021}. On the other hand, acquiring 
information experimentally or theoretically is costly too. At the end, the second law (and thermodynamics in general) 
is about a tradeoff between knowing the essential features \emph{and} retaining an efficient description. Thus, the 
question which information is available and which information renders the description efficient, depends---at the very 
end---on the particular situation. As long as the bath is not too small (say, not less than 100 qubits), we believe 
that the description given in the main text is the most adequate one for many purposes. 

%%%%%%%%%%%%%%%%%%%%%%%%%%%%%%%%%%%%%%%%%%%%%%%%%%%%%%%%%%%%%%%%%%%%%%%%%%%%%%%%%%%%%%%%%%%%%%%%%%%%%%%%%%%%%%%%%%%%%%%%

\end{document}